\documentstyle[12pt,epsf]{article}

\title{Front motion for phase transitions in systems with memory}
\date{}

\author{\\
{\bf Horacio G. Rotstein\(^{1}\)}
\thanks{Current address: Brandeis University, Dept. of Chemistry and Volen 
Center for Complex Systems, MS 015, Waltham, MA, 02454-9110, USA. 
E-mail: horacio@cs.brandeis.edu},
{\bf Alexander I. Domoshnitsky\(^{3}\)}, \\
{\bf Alexander Nepomnyashchy\(^{1,2}\)},
\\ 
\(^{1}\) Department of Mathematics and \\
\(^{2}\) Minerva Center for Nonlinear Physics of Complex Systems, \\
Technion - IIT, Haifa, 32000, Israel \\  
\(^{3}\) The Research Institute, The College of Judea and Samaria, \\
Ariel 44837, Israel \\ \\
}

\def\alf{\alpha}
\def\half{\bar{\alf}}
\def\ihalf{\bar{\alf_{i}}}\def\gam{\gamma}
\def\kap{\kappa}
 
\def\bet{\beta}
\def\oalf{\alpha_{1}}
\def\ohalf{\bar{\oalf}}
\def\ogam{\gamma_{1}}
\def\talf{\alpha_{2}}
\def\thalf{\bar{\talf}}
\def\tgam{\gamma_{2}}
\def\epsl{\epsilon}
\def\eps2{\epsilon^{2}}
\def\eln3{\epsilon^{3}}
\def\ep4s{\epsilon^{4}}
\def\invepsl{\frac{1}{\epsilon}}
\def\inveps2{\frac{1}{\epsilon^{2}}}
\def\inv3eps{\frac{1}{\epsilon^{3}}}
\def\phit{\phi_{t}}
\def\phitt{\phi_{tt}}

\def\phixx{\phi_{xx}}

\def\phiyy{\phi_{yy}}
\def\phitxx{\phi_{xxt}}
\def\phityy{\phi_{yyt}}

\def\zphi{\phi^{0}}
\def\ophi{\phi^{1}}
\def\tphi{\phi^{2}}
\def\Phit{\Phi_{t}}
\def\Phiz{\Phi_{z}}

\def\Phitt{\Phi_{tt}}
\def\Phixx{\Phi_{xx}}
\def\Phizz{\Phi_{zz}}
\def\Phizt{\Phi_{zt}}
\def\Phizx{\Phi_{zx}}

\def\Phitxx{\Phi_{xxt}}
\def\Phizxx{\Phi_{xxz}}
\def\Phitxz{\Phi_{xzt}}
\def\Phixzz{\Phi_{xzz}}
\def\Phitzz{\Phi_{zzt}}
\def\Phizzz{\Phi_{zzz}}
\def\Phir{\Phi_{r}}
\def\Phis{\Phi_{s}}
\def\Phirr{\Phi_{rr}}
\def\Phiss{\Phi_{ss}}
\def\Phirt{\Phi_{rt}}
\def\Phirs{\Phi_{rs}}
\def\Phist{\Phi_{st}}

\def\Phizs{\Phi_{zs}}
\def\zPhi{\Phi^{0}}
   
\def\zPhiz{\Phi_{z}^{0}}

\def\zPhizz{\Phi_{zz}^{0}}
\def\zPhizt{\Phi_{zt}^{0}}
\def\zPhizx{\Phi_{zx}^{0}}

\def\zPhixi{\Phi_{\xi}^{0}}
\def\zPhixixi{\Phi_{\xi\xi}^{0}}

\def\zPhizs{\Phi_{zs}^{0}}

\def\oPhi{\Phi^{1}}
\def\tPhi{\Phi^{2}}

\def\oPhizz{\Phi_{zz}^{1}}

\def\oPhixixi{\Phi_{\xi\xi}^{1}}

\def\Up{\Upsilon}
\def\Upt{\Upsilon_{t}}
\def\Upz{\Upsilon_{z}}

\def\Ups{\Upsilon_{s}}

\def\zUp{\Upsilon^{0}}
\def\zUpt{\Upsilon_{t}^{0}}   
\def\zUpz{\Upsilon_{z}^{0}}

\def\zUps{\Upsilon_{s}^{0}}

\def\oUp{\Upsilon^{1}}
\def\oUpt{\Upsilon_{t}^{1}}   
\def\oUpz{\Upsilon_{z}^{1}}

\def\oUps{\Upsilon_{s}^{1}}

\def\tUpz{\Upsilon_{z}^{2}}

\def\ut{u_{t}}
\def\utt{u_{tt}}

\def\Uz{U_{z}}

\def\Us{U_{s}}

\def\Uss{U_{ss}}

\def\Ust{U_{st}}
\def\Utt{U_{tt}}
\def\Uzz{U_{zz}}
\def\Uzt{U_{zt}}
\def\Uzs{U_{zs}}
\def\zu{u^{0}}
\def\ou{u^{1}}

\def\zU{U^{0}}
\def\oU{U^{1}}

\def\zut{u_{t}^{0}}   
   
\def\zlapu{\Delta u^{0}}
\def\out{u_{t}^{1}}   
   
\def\olapu{\Delta u^{1}}
   
\def\zUz{U_{z}^{0}}

\def\zUs{U_{s}^{0}}
\def\zUss{U_{ss}^{0}}
\def\zUst{U_{st}^{0}}
\def\zUtt{U_{tt}^{0}}
\def\zUzz{U_{zz}^{0}}
\def\zUzt{U_{zt}^{0}}
\def\zUzs{U_{zs}^{0}}
   
\def\oUz{U_{z}^{1}}

\def\oUzz{U_{zz}^{1}}
\def\oUzt{U_{zt}^{1}}
\def\tUzz{U_{zz}^{2}}

\def\vt{v_{t}}

\def\Vt{V_{t}}
\def\Vz{V_{z}}

\def\Vs{V_{s}}

\def\zV{V^{0}}

\def\zVt{V_{t}^{0}}   
\def\zVz{V_{z}^{0}}

\def\zVs{V_{s}^{0}}

\def\oVz{V_{z}^{1}}

\def\est{S_{t}}
\def\st2{S_{t}^{2}}
\def\esx{S_{x}}
\def\sx2{S_{x}^{2}}
\def\stt{S_{tt}}
\def\sxx{S_{xx}}

\def\tsxx{S_{xxx}}
\def\sxt{S_{xt}}

\def\zst2{S_{0,t}^{2}}

\def\zsx2{S_{0,x}^{2}}

\def\ost2{S_{1,t}^{2}}

\def\osx2{S_{1,x}^{2}}

\def\xix{\xi_{x}}
\def\xiz{\xi_{z}}
\def\xit{\xi_{t}}
\def\xizx{\xi_{zx}}
\def\xizt{\xi_{zt}}

\def\rht2{\rho_{t}^{2}}

\def\rhc2{\rho_{\theta}^{2}}

\def\zrht{\rho_{t}^{0}} 
\def\zrht2{\rho_{t}^{2}^{0}}
\def\zrhc{\rho_{\theta}^{0}}
\def\zrhc2{\rho_{\theta}^{2}^{0}}

\def\orht{\rho_{t}^{1}}
\def\orht2{\rho_{t}^{2}^{1}}
\def\orhx{\rho_{\theta}^{1}}
\def\orhx2{\rho_{\theta}^{2}^{1}}

\def\xix{\xi_{x}}
\def\xiz{\xi_{z}}
\def\xit{\xi_{t}}
\def\xizx{\xi_{zx}}
\def\xizt{\xi_{zt}}

\def\ar2{R_{0}^{2}}

\def\art2{R_{0,t}^{2}}

\def\er2{R^{2}}

\def\ert2{R_{t}^{2}}

\def\ardt{r_{t}}
\def\radt2{r_{t}^{2}}
\def\radtt{r_{tt}}
\def\ract{s_{t}}
\def\arct2{s_{t}^{2}}
\def\arctt{s_{tt}}

\def\vel2{v^{2}}
\def\u-lap{\Delta u}
\def\lap-phi{\Delta \phi}
\def\lapt-phi{\Delta \phit}

\def\laprad{\Delta r}
\def\laparc{\Delta s}

\def\gradrad2{\left| \nabla r \right| ^{2}}
\def\gradarc2{\left| \nabla s \right| ^{2}}

\def\xicart{1 + \sx2 - \talf\ \st2} 
\def\nxicart{1 + \sx2 - \alf\ \st2} 
\def\axicart{(\xicart)^{\frac{1}{2}}}
\def\bxicart{(\xicart)^{\frac{3}{2}}}
\def\nbxicart{(\nxicart)^{\frac{3}{2}}}
\def\naxicart{(\nxicart)^{\frac{1}{2}}}

\def\invaxicart{(\xicart)^{-\frac{1}{2}}}
\def\invbxicart{(\xicart)^{-\frac{3}{2}}}

\def\dximove{1 - \talf\ \radt2} 
\def\adximove{(\dximove)^{\frac{1}{2}}} 
\def\bdximove{(\dximove)^{\frac{3}{2}}} 

\def\tximove{1 -  \alf\ \radt2} 
\def\taximove{(\tximove)^{\frac{1}{2}}} 
\def\tbximove{(\tximove)^{\frac{3}{2}}}

\def\tinvaximove{(\tximove)^{-\frac{1}{2}}} 
\def\tinvbximove{(\tximove)^{-\frac{3}{2}}}

\def\t*{t_{*}}
\def\inoi{\int_{0}^{\infty}\ }
\def\inii{\int_{-\infty}^{\infty}\ }
\def\begeq{\begin{equation}}
\def\endeq{\end{equation}}
\def\begdis{\begin{displaymath}}
\def\enddis{\end{displaymath}}
\def\bk{\bigskip}

\def\nd{\noindent}

\begin{document}

\maketitle

\begin{abstract}

\nd We consider the following partial integro-differential equation 
(Allen-Cahn equation with memory)

\begdis
	\eps2\ \phit = \int_{0}^{t} a(t-t')\ [ \eps2\ \lap-phi + f(\phi)\ +
	\epsl\ h] (t')
	dt'.					   
\enddis

\nd where \( \epsl \) is a small parameter, \( h \) is a constant, 
\( f(\phi) \) is minus the derivative of a double well potential and the 
kernel \( a \) is a piecewise continuous, differentiable at the origin, 
scalar-valued function on \( (0,\infty) \). The prototype kernels are 
exponentially decreasing functions of time and they reduce the 
integrodifferential equation to a hyperbolic one, the damped Klein-Gordon
equation. By means of a formal asymptotic analysis we show that to the leading 
order and under suitable 
assumptions on the kernels, the integro-differential equation behave like a 
hyperbolic partial differential equation obtained by considering prototype 
kernels: the evolution of fronts is governed by the extended, damped 
Born-Infeld equation.
\bk

\nd We also apply our method to a system of partial integro-differential 
equations which generalize the classical phase field equations with a 
non-conserved order parameter and describe the process of phase transitions 
where memory effects are present, 

\begdis
	\left\{ \begin{array}{l}
		u_{t} + \eps2\ \phi_{t} = \int_{0}^{t} a_{1}(t-t')\ 
		\u-lap(t') dt'   \\
		                             			    \\
		\eps2\ \phi_{t} = \int_{0}^{t} a_{2}(t-t')\ 
		[\ \eps2\ \lap-phi + f(\phi) + \epsl\ u\ ](t') 
		dt',
		\end{array}
	\right.
\enddis
	
\nd where \( \epsl \) is a small parameter. In this case the functions \( u \) 
and \( \phi \) represent the temperature field and order parameter 
respectively. The kernels \( a_{1} \) and \( a_{2} \) are assumed to be similar
to \( a \). For the phase field equations with memory we obtain the same
result as for the generalized Klein-Gordon equation or Allen-Cahn equation 
with memory.

\end{abstract} 

\section{Introduction}

\nd In this manuscript we consider the following partial 
integro-differential equation

\begeq
	\eps2\ \phit = \int_{0}^{t} a(t-t')\ [ \eps2\ \lap-phi + f(\phi)\ +
	\epsl\ h] (t')\ dt',					   
						\label{eq:memphasefield22}
\endeq

\nd where \( \phi(x,t) \) is a field, called an order parameter, defined in 
\( \Omega \times [0,T] \), where \( \Omega \subset R^{2} \) and \( T > 0 \), 
with Dirichlet and Neumann boundary conditions. 
The parameter \( \epsl \) is assumed to be small, 
\( \epsl \ll 1 \), \( f(\phi) \) is a real odd function with a positive 
maximum equal to \( \phi^{\ast} \), a negative minimum equal to 
\( -\phi^{\ast} \) and precisely three zeros in the closed interval 
\( [-b,b] \) located at \( 0 \) and \( \pm b \), where \( b \) is a positive 
constant. For simplicity we will consider 
\( b = 1 \). The operator \( \Delta \) is the Laplace operator. The kernel 
\( a \) is assumed to be a  piecewise continuous, 
differentiable at the origin and scalar-valued functions on 
\( (0,\infty) \) satisfying additional conditions to be described later.
\bk

\nd Specific cases of equation (\ref{eq:memphasefield22}) are 
\( a(t) = \delta(t) \), the Allen-Cahn equation:

\begeq
	\eps2 \phit = \eps2\ \lap-phi + f(\phi)\ + \epsl\ h,
						\label{eq:allen-cahn}
\endeq

\nd which is the simplest model of phase transition with a non-conserved order
parameter, and  the damped Klein-Gordon equation (used, e.g., in the theory of
long Josephson junctions):

\begeq
	\eps2\ \phitt + \gam\ \eps2\ \phit = \eps2\ \lap-phi + f(\phi)
	+ \epsl\ h,
							\label{eq:phasefield22}
\endeq

\nd which is obtained for an exponentially decreasing kernel of type
\cite{kn:rotnep1}

\begeq
	a(t) = e^{-\gam\ t}.				\label{eq:kernel}
\endeq

\nd Equation (\ref{eq:memphasefield22}) can be thought of as a phenomenological
equation based on energetic penalization driving the evolution of the system 
toward equilibrium states. More specifically, let us call
\( F_{\epsl} = \int_{R^{2}} [\frac{1}{2} \eps2 (\nabla \phi)^{2} - V(\phi)] 
d\bar{x} \), with \( f(\phi) + \epsl\ h = dV / d\phi \) the free energy of
the system. The functional derivative of the free energy, 
\( \delta F_{\epsl}(\phi) / \delta \phi \)
is considered as a generalized force indicative of the tendency of the free 
energy to decay towards a minimum. The equation is obtained by assuming  that 
the response of \( \phi \) to the tendency of the free energy to decay towards 
a minimum is given by \cite{kn:rotnep1,kn:rot1}, 

\begdis
	\eps2 \phit = - \int_{0}^{t} a(t-t')\ 
	\frac{\delta F_{\epsl}}{\delta \phi}\ (t')\ dt'. 
\enddis

\nd A system of integro-differential equations closely related to 
(\ref{eq:memphasefield22}), the phase field equations with memory, is
\cite{kn:rotnep3,kn:rot1}

\begeq
	\left\{ \begin{array}{l}
		u_{t} + \eps2\ \phi_{t} = \int_{0}^{t} a_{1}(t-t')\ 
		\u-lap(t') dt'\\
		                             \label{eq:defas2}    \\
		\eps2\ \phi_{t} = \int_{0}^{t} a_{2}(t-t' 
		[\ \eps2\ \lap-phi + f(\phi) + \epsl\ u\ ](t')
		dt'
		\end{array}
	\right.
\endeq
\bk

\nd in \( \Omega \times [0,t] \) where \( \Omega \subset R^{2} \) and 
\( T > 0 \) with Dirichlet and Neumann 
boundary conditions for the temperature field, \( u(x,t) \) and the order 
parameter, \( \phi(x,t) \), respectively and for \( u \) vanishing initially.
The parameter \( \epsl \) and \( f(\phi) \) are as for 
(\ref{eq:memphasefield22}) and the kernels \( a_{1} \) and \( a_{2} \) are
similar to \( a \); i.e., piecewise continuous, differentiable at the origin, and scalar-valued functions on \( (0,\infty) \). 
\bk

\nd For \( a_{1}(t) = a_{2}(t) = \delta(t) \), system (\ref{eq:defas2}) 
gives rise to the classical phase field equations with a non-conserved order 
parameter

\begeq
	\left\{\begin{array}{l}
		\ut + \eps2 \phit = \u-lap,				\\
									\\
		\eps2 \phit = \eps2 \lap-phi + f(\phi) + \epsl\ u.
		\end{array}
	\right.
						\label{eq:defas3}
\endeq
\bk

\nd System (\ref{eq:defas2}) also generalizes the hyperbolic phase field 
equations with a non-conserved order parameter:

\begeq
	\left\{\begin{array}{l}
		\utt + \eps2\ \phitt + \ogam\ \ut + \eps2\ \ogam\ \phit =
	        \alf\ \u-lap,				\\
						\label{eq:defas1} \\
		\eps2\ \phitt + \eps2\ \tgam\ \phit = 
		\eps2\ \lap-phi + f(\phi) + \epsl\ u,
		\end{array}
	\right.
\endeq

\nd reducing to them when the kernels \( a_{1} \) and \( a_{2} \) are 
exponentially decreasing functions; i.e., \( a_{i}(t) = \alf_{i}\ 
e^{-\gam_{i}\ t} \), \( i = 1, 2 \), with \( \alf_{i} \) and \( \gam_{i} \) 
are non-negative constants. System (\ref{eq:defas1}) is obtained by 
differentiating both equations in (\ref{eq:defas2}), rearranging terms and 
rescaling by means of the transformation
 \( t \rightarrow \talf^{-\frac{1}{2}}\ t \), 
\( \tgam \rightarrow \talf^{\frac{1}{2}}\ \tgam \), 
\( \ogam \rightarrow \talf^{\frac{1}{2}}\ \ogam \) and
\( \alf := \oalf / \talf \). 
\bk

\nd The phase field equations with memory (\ref{eq:defas2}) describe the 
process of phase transitions where memory effects are present. For a readable
description of the classical phase field equations see \cite{kn:cag1}. The 
first equation in (\ref{eq:defas2}) is based on the balance of heat equation 
for a non-Fourier 
process in which the expression for the heat flux is given by a convolution
in time between the temperature gradient and the kernel \( a_{1} \)
\cite{kn:aizbar1,kn:nov1}. The second equation is obtained in a similar way as
equation (\ref{eq:memphasefield22}) \cite{kn:rot1}. 
\bk

\nd In what follows we assume that there exists a solution \( \phi(x,t) \) 
of (\ref{eq:memphasefield22}) or \( \{ u(x,t), \phi(x,t) \} \) of 
(\ref{eq:defas2}) defined for all small \( \epsl \), every \( x \in \Omega \) 
and every \( t \in [0,T] \) which contains an internal layer. We also 
assume, for such solutions, that for all small \( \epsilon \geq 0 \) and all 
\( t \in [0,T] \), the domain \( \Omega \) can be divided into two  
open regions \( \Omega_{+}(t;\epsilon) \) and \( \Omega_{-}(t,\epsilon) \) 
with a curve \( \Gamma(t;\epsilon) \), 
separating between them. This interface defined by 

\begeq
	\Gamma(t;\epsilon) := \left\{ x \in \Omega : \phi(x,t;\epsilon) = 0 
	\right\}, 
							\label{definterf}
\endeq

\nd is assumed to be smooth, which implies that its curvature and its 
velocity are bounded independently of \( \epsilon \). The function \( \phi \)
is assumed to vary continuously across the interface, far from the interface
tending to \( 1 \) when \( x \in \Omega_{+}(t;\epsilon) \), -1 when 
\( x \in \Omega_{-}(t,\epsilon) \), with rapid spatial variation close to
the interface. The problem is to derive a closed equation for the evolution of 
the interface asymptotically valid as \( \epsl \rightarrow 0 \).
\bk

\nd The latter problem has been studied formerly by means of formal asymptotic
analysis for PDEs (\ref{eq:allen-cahn}) and (\ref{eq:phasefield22})
\cite{kn:cagfif1,kn:fif1,kn:fifpen1}, as well as for systems (\ref{eq:defas3})
and (\ref{eq:defas1}) \cite{kn:rotnep1,kn:neu1}. For 
(\ref{eq:allen-cahn}) and (\ref{eq:defas3}), fronts evolve according to the 
mean curvature flow equation \cite{kn:rubste1,kn:cag3,kn:cag2}, 

\begeq
	v = \kap,					\label{eq:flowmc}
\endeq 

\nd where \( v \) is the normal velocity of the interface and \( \kap \) its
curvature. If \( y = S(x,t) \) is the Cartesian description of the interface, 
equation (\ref{eq:flowmc}) reads

\begdis
	\est = \frac{\sxx}{1 + \sx2}.
\enddis

\nd For the undamped Klein-Gordon equation, (\ref{eq:phasefield22}) with 
\( \gam = h = 0 \), Neu \cite{kn:neu1} proved that the evolution of the 
interface is governed by the Born-Infeld equation

\begeq
	( 1 - \st2 )\ \sxx + 2\ \esx\ \est\ \sxt - ( 1 + \sx2 )\ \stt
	= 0.						\label{eq:bicart}
\endeq

\nd For (\ref{eq:phasefield22}) with \( \gam > 0 \) and for (\ref{eq:defas1}) 
Rotstein et. al. \cite{kn:rotnep1,kn:rot1} showed that fronts move 
according to an extended version of the Born-Infeld equation given by

\begdis
	( 1 - \alf\ \st2 )\ \sxx + 2\ \alf\ \esx\ \est\ \sxt\ - 
	\alf\ ( 1 + \sx2 )\ \stt -
\enddis

\begeq
	- \gam\ \est\ ( \nxicart )  - \nu\ \nbxicart = 0,   
							\label{eq:mbornexcart}
\endeq

\nd \nd where the parameter \( \nu \) (proportional to \( h \) in 
(\ref{eq:phasefield22})) is defined later. 
\bk

\nd In local (geometric) coordinates equation (\ref{eq:mbornexcart}) reads

\begeq
	\frac{v_{t}}{1 - \alf\ v^{2}} + \gam\ v = \kap + \nu\
	(1 - \alf\ v^{2})^{\frac{1}{2}}.	
							\label{eq:flowmcmemph}
\endeq

\nd where \( v \), \( v_{t} \) and \( \kap \) are the normal velocity, normal 
acceleration  and curvature of the interface respectively. Equation 
(\ref{eq:mbornexcart}) or (\ref{eq:flowmcmemph}) have been studied in 
\cite{kn:rot1}, and in \cite{kn:neu1} for \( \gam = \nu = 0 \)  .
\bk

\nd The case of integrodifferential equations (\ref{eq:memphasefield22}) and 
(\ref{eq:defas2}) is still less investigated.
\nd For (\ref{eq:defas2}) and for kernels which are Laplace transforms of 
suitable functions satisfying

\begdis
	\int_{0}^{\infty} a(t)\ dt < \infty, \ \ \ \ \ \ \ \ \ \ 
	\int_{0}^{\infty} \bar\alf(s)\ ds < \infty \ \ \ \ \ \ \
	\mbox{and} \ \ \ \ \ \ \ 
	\int_{0}^{\infty} \bar\alf(s)\ s\ ds < \infty 
\enddis

\nd where \( \bar\alf(s) \) is the inverse Laplace transform of \( a(t) \),
Rotstein et al. \cite{kn:rot1,kn:rotnep5} showed that the equation governing 
the motion of the interface is also (\ref{eq:mbornexcart}) where in this case 

\begdis
	\gam = - \frac{a'_{2}(0)}{a_{2}(0)^{2}}.
\enddis

\nd In the present manuscript we show that in the asymptotic limit studied and
for a certain class of kernels, which will be made clear below, equation 
(\ref{eq:memphasefield22}) is equivalent to (\ref{eq:phasefield22}) from the
point of view of fronts motion; i.e.,
the evolution of interfaces is described by the same equation,
(\ref{eq:mbornexcart})  
with

\begdis
 	\alf = \left( \inoi \half(\tau)\ d\tau \right)^{-1},
\enddis

\nd where \( \half(\tau) \) is the inverse Laplace transform of 
\( a(t) \), and

\begdis 
 	\gam = \alf^{2}\ \left( \inoi \half(\tau)\ \tau\ d\tau \right).
\enddis

\nd We also find the equivalence between (\ref{eq:defas2}) and
(\ref{eq:defas1}) \cite{kn:rotnep5} for a more general class of kernels,
described below.  
In Section 2 we describe the class of kernels considered in this work,
give some examples and explain our strategy in dealing with them. 
In Section 3 we derive the equation of motion for (\ref{eq:memphasefield22}). 
We present the derivation treating the equation in Cartesian coordinates and
representing the kernel as a Laplace tranform
of a suitable function. For kernels represented as a Fourier transform of 
suitable functions the derivation is similar. In our derivation we assume that
the interface has no oscillations in the sense that there are no points on the
interface for which the velocity vanishes; i.e., the front is an advancing 
front. This assumption is not necessary for the case of exponentially 
decreasing kernels; i.e., for (\ref{eq:phasefield22}). In Section 4 we derive
the equation of motion for (\ref{eq:defas2}) for kernels which are inverse
Fourier transforms of suitable functions. Here we treat the problem in local
coordinates.

\section{Description of the class of kernels considered}

\nd In order to deal with the memory integral in deriving the equations of 
front motion for either (\ref{eq:memphasefield22}) or (\ref{eq:defas2}) we 
represent the kernels \( a(t) \) and \( a_{i}(t) \), \( i = 1, 2 \), as 
Laplace transforms of suitable functions:

\begeq
	a(t) = \inoi \half(\tau)\ e^{-\tau\ t}\ d\tau,
							\label{eq:kernel-lap1}
\endeq

\nd or

\begeq
	a_{i}(t) = \inoi \ihalf(\tau)\ e^{-\tau\ t}\ d\tau.
							\label{eq:kernel-lap2}
\endeq

\nd for \( i = 1, 2 \). Substituting (\ref{eq:kernel-lap1}) into 
(\ref{eq:memphasefield22}), rearranging terms and defining

\begeq
	\chi(t;\tau)  = \int_{0}^{t} e^{-\tau\ (t-t')}\ 
	[\ \eps2\ \lap-phi + f(\phi)\ + \epsl\ h\ ] (t')\ dt',
\endeq 	

\nd where it is understood that \( \chi \) is also a function of the space 
variable \( x \) and it is defined for all \( x \in \Omega \), we obtain

\begeq
	\left\{ 
		\begin{array}{l}	
		\eps2\ \phit = \inoi \half(\tau)\ \chi(t;\tau)\ d\tau, \\
						\label{eq:memphasefield2comp}\\
		\chi_{t}(t;\tau) + \tau\ \chi(t;\tau) = \eps2\ \lap-phi + 
		f(\phi)\ + \epsl\ h,
		\end{array}
	\right.
\endeq

\nd for all \( \tau \in [0,\infty) \). Note that the second equation in 
(\ref{eq:memphasefield2comp}) satisfies \( \chi(0,\tau) = 0 \) for all 
\( x \in \Omega \) and for all
\( \tau \in [0,\infty) \). We can see that, 
in spite of the fact that there is an integral involved 
in the first equation, that integral sums over the parameter \( \tau \) not
involving the time variable \( t \) as in (\ref{eq:memphasefield22}). 
For the phase field equations with memory, substituting 
(\ref{eq:kernel-lap2}) into (\ref{eq:defas2}), rearranging 
terms and calling

\begeq
	\chi(t;\tau)  = \int_{0}^{t} e^{-\tau\ (t-t')}\ 
	[\ \eps2\ \lap-phi + f(\phi)\ + \epsl\ u\ ] (t')\ dt',
\endeq 	

\nd and

\begeq
	v(t;\tau) = \int_{0}^{t} e^{-\tau\ (t-t')}\ \u-lap  (t')\ dt',
\endeq

\nd where, again, it is understood that \( \chi \) and \( v \) are also 
functions of the space variable \( x \) and are defined for all \( x \in 
\Omega \), we obtain

\begeq
	\left\{ 
		\begin{array}{l}	
		\ut + \eps2\ \phit = \inoi \ohalf(\tau)\ v(t;\tau) d\tau,   \\
									\\
		\vt(t;\tau) + \tau\ v(t;\tau) = \u-lap,				  	\\
									\\
		\eps2\ \phit = \inoi \half(\tau)\ \chi(t;\tau)\ d\tau, 	\\
									\\
		\chi_{t}(t;\tau) + \tau\ \chi(t;\tau) = \eps2\ \lap-phi + 
		f(\phi) + \epsl\ u,			\label{eq:defas2comp}
		\end{array}
	\right.
\endeq

\nd for all \( \tau \in [0,\infty) \). Note that the second and fourth 
equations satisfy \( v(0,\tau) = \chi(0,\tau) = 0 \) for all 
\( x \in \Omega \) and for all \( \tau \in [0,\infty) \). 
\bk

\nd For an exponentially decreasing kernel (\ref{eq:kernel}), 
\( \half(\tau) = \delta(\tau - \gam) \). For a kernel which is a linear 
combination of kernels of type (\ref{eq:kernel}), 
\( a(t) = \sum_{k=1}^{n} \alf_{k}\, e^{-\gam_{k} t} \), for positive constants
\( \alf_{k} \) and \( \gam_{k} \) (\( k = 1, 2, \ldots, n \)), we have
\( \half(\tau) = \sum_{k=1}^{n} \alf_{k}\, \delta(\tau - \gam_{k}) \). In the
latter case we can define for \( k = 1, 2, \ldots, n \)

\begeq
	\chi(t;\gam_{k})  = \int_{0}^{t} e^{-\gam_{k}\ (t-t')}\ 
	[\ \eps2\ \lap-phi + f(\phi)\ + \epsl\ h\ ] (t')\ dt',
\endeq 	

\nd and substitute into (\ref{eq:memphasefield22}) obtaining

\begeq
	\left\{ 
		\begin{array}{l}	
		\eps2\ \phit = \sum_{k=1}^{n} \alf_{k}\ \chi(t;\gam_{k}), \\
						\label{eq:memphasefield2sum}\\
		\chi_{t}(t;\gam_{k}) + \gam_{k}\ \chi(t;\gam_{k}) = 
		\eps2\ \lap-phi + f(\phi)\ + \epsl\ h,
		\end{array}
	\right.
\endeq

\nd for \( k = 1, 2, \ldots, n \). It is clear that 
(\ref{eq:memphasefield2sum}) is a particular (discrete) case of 
(\ref{eq:memphasefield2comp}) and the simplest generalization of 
(\ref{eq:phasefield22}). This discussion can be easily adapted to 
system (\ref{eq:defas2}). 
\bk

\nd The approach discussed up to now in this section allows us to consider 
kernels which are not necessarily exponentially decreasing. Examples are, 
for positive \( \alf \), \( \bet \) and \( \gam \): 
\( a(t) = 1 / (\gam t + \alf) \), for which \( \half(\tau) = 
1 / \gam\, e^{-\frac{\alf}{\gam}t} \) or, more generally
\( a(t) = 1 / (\gam t + \alf)^{n} \) (with \( n \) a positive integer), 
for which \( \half(\tau) = 1 / (\gam^{n}\ (n-1)! )\ \tau^{n-1}\ 
e^{-\frac{\alf}{\gam}t} \); \( a(t) = \bet / ((\gam\ t + \alf)^{2} + 
\bet^{2}) \) and \( a(t) = (\gam t + \alf)  / ((\gam t + \alf)^{2} + 
\bet^{2}) \) for which \( \half(\tau) = 1 / \gam^{2}\ 
e^{-\frac{\alf}{\gam}t} \sin(\bet\ \tau) \) and \( \half(\tau) = 
1 / \gam^{2}\ e^{-\frac{\alf}{\gam}t} \cos(\bet\ \tau) \) respectively.
A class of kernels not included in (\ref{eq:kernel-lap1}) or 
(\ref{eq:kernel-lap2}) is \( a(t) = \alf\ e^{-\gam\ t} \cos(\bet\ t) \), or
equivalently, \( a(t) = \alf\ \frac{e^{-z_{1}\ t} + e^{-\bar{z}_{1} t}}{2} 
\) where \( z_{1} = \gam - i\ \bet \). 
\bk

\nd From a slightly different point of view, kernels of type
\( a(t) = \alf\ e^{-\gam\ t} \cos(\bet\ t) \) can be represented as Fourier 
transforms of suitable functions.

\begeq
	a(t) = \inii \half(w)\ e^{-i w t}\ dw,
							\label{eq:kernel-fou1}
\endeq

\nd or

\begeq
	a_{i}(t) = \inii \ihalf(w)\ e^{-i w t}\ dw.
							\label{eq:kernel-fou2}
\endeq

\nd for \( i = 1, 2 \) where 

\begdis
	\half(w) = \frac{1}{2 \pi}\ \inii a(t)\ e^{i w t}\ dt,
\enddis

\nd and

\begdis
	\ihalf(w) = \frac{1}{2 \pi}\ \inii \ a_{i}(t)\ e^{i w t}\ dt.
\enddis

\nd for \( i = 1, 2 \). In this case the functions \( a(t) \) and 
\( a_{i}(t) \), \( i =1, 2 \) are understood to be defined for all \( t \) and
being equal to zero for \( t < 0 \). Substituting (\ref{eq:kernel-fou1}) into 
(\ref{eq:memphasefield22}), rearranging terms and defining

\begeq
	\chi(t;w)  = \int_{0}^{t} e^{-i w (t-t')}\ 
	[\ \eps2\ \lap-phi + f(\phi)\ + \epsl\ h\ ] (t')\ dt',
\endeq 	

\nd where it is understood that \( \chi \) is a function of the space variable
\( x \in \Omega \), we obtain

\begeq
	\left\{ 
		\begin{array}{l}	
		\eps2\ \phit = \inii \half(w)\ \chi(t;w)\ dw, \\
						\label{eq:memphasefield2fou}\\
		\chi_{t}(t;w) + i\ w\ \chi(t;w) = \eps2\ \lap-phi + 
		f(\phi)\ + \epsl\ h,
		\end{array}
	\right.
\endeq

\nd for all \( w \in (-\infty,\infty) \). For the phase field equations with 
memory, substituting (\ref{eq:kernel-fou2}) into (\ref{eq:defas2}), rearranging
terms and calling

\begeq
	\chi(t;w)  = \int_{0}^{t} e^{-i w (t-t')}\ 
	[\ \eps2\ \lap-phi + f(\phi)\ + \epsl\ u\ ] (t')\ dt',
\endeq 	

\nd and

\begeq
	v(t;w) = \int_{0}^{t} e^{-i w (t-t')}\ \u-lap  (t')\ dt',
\endeq

\nd where, again, it is understood that \( \chi \) and \( v \) are functions
of the space variable \( x \in \Omega \). System (\ref{eq:defas2}) becomes, 
for all \( w \)

\begeq
	\left\{ 
		\begin{array}{l}	
		\ut + \eps2\ \phit = \inii \ohalf(w)\ v(t;w) dw,   \\
									\\
		\vt(t;w) + i\ w\ v(t;w) = \u-lap,				  	\\
									\\
		\eps2\ \phit = \inii \half(w)\ \chi(t;w)\ dw, 	\\
									\\
		\chi_{t}(t;w) + i w chi(t;w) = \eps2\ \lap-phi + 
		f(\phi) + \epsl\ u,			\label{eq:defas2fou}
		\end{array}
	\right.
\endeq

\nd for all \( w \in (-\infty,\infty) \). As we set at the beginning of the 
introduction, the kernels \( a(t) \), are assumed to be 
piecewise continuous, differentiable at the origin, scalar-valued
functions on \( (0,\infty) \). Moreover they are assumed to be independent of 
\( \epsl \) and such that 

\begdis
	\int_{0}^{\infty} a(t)\ dt < \infty, 
\enddis

\nd and

\begdis
	\int_{0}^{\infty} \bar\alf(s)\ ds < \infty \ \ \ \ \ \ \
	\mbox{and} \ \ \ \ \ \ \
	\int_{0}^{\infty} \bar\alf(s)\ s\ ds < \infty. 
\enddis

\nd when we use the Laplace transform, and 

\begdis
	\int_{-\infty}^{\infty} \bar\alf(s)\ ds < \infty \ \ \ \ \ \ \
	\mbox{and} \ \ \ \ \ \ \
	\int_{-\infty}^{\infty} \bar\alf(s)\ s\ ds < \infty. 
\enddis

\nd when we use the Fourier transform. In both cases this is equivalent to 
\( a(0) < \infty \) and \( a'(0) < \infty \).

\section{The Allen-Cahn equation with memory - asymptotic analysis in 
Cartesian coordinates}

\nd For points outside the interface and for \( \epsl > 0 \) we expand 
\( \phi \) as follows

\begdis
	\phi = \phi(x,t;\epsl) = \zphi(x,t) + \epsl\ \ophi(x,t) + 
	\eps2\ \tphi(x,t) + {\cal O}(\epsl^{3}),
\enddis

\nd and substitute into (\ref{eq:memphasefield22}) obtaining the 
\( {\cal O}(1) \) and \( {\cal O}(\epsl) \) respectively

\begdis
	\int_{0}^{t} a(t-t')\ f(\phi^{0})\ dt' = 0,
\enddis

\nd whose solution is \( f(\phi^{0}) = 0 \) (or \( \phi^{0} = \pm\ 1 \)), and

\begdis
	\int_{0}^{t} a(t-t')\ [ f'(\phi^{0})\ \phi^{1} + h ]\ dt' = 0,
\enddis

\nd giving \( f'(\phi^{0})\ \phi^{1} + h = 0 \). Thus, for points which have
not yet been reached by the moving front, we have 

\begeq
	\chi^{0}(t;\tau) = 0,			\label{eq:outsideint0}
\endeq

\nd and

\begeq
	\chi^{1}(t;\tau) = 0,			\label{eq:outsideint1}
\endeq

\nd for all \( \tau \in [0,\infty) \).
\bk

\nd For the asymptotic analysis using Cartesian coordinates the interface is 
represented by 

\nd \( y = S(x,t,\epsl) \) for \( \epsl \) sufficiently small and 
assume \( \est \neq 0 \) for all \( x \in \Omega \) and all \( t \geq 0 \). We
define a new variable

\begdis
	z := \frac{y - S(x,t,\epsl)}{\epsl}	
\enddis

\nd which is assumed to be \( {\cal O}(1) \) as \( \epsl \rightarrow 0 \) 
near the interface. We
call \( \Phi \) the asymptotic form of \( \phi \) as \( \epsl \rightarrow 0 \)
with \( z \) fixed; i.e.,  

\begeq	
	\phi = \Phi(z,x,t;\epsl).			\label{eq:cartint}
\endeq

\nd The field equations (\ref{eq:memphasefield2comp})  in \( (z,x,t) \)
coordinates, after differentiating the first equation with respect to \( t \)
and calling \( \chi = \Up(x,z,t;\tau,\epsl) \),
become (see Appendix A)

\begeq
	\left\{ 
		\begin{array}{l}
		\eln3\ \Phitt - 2\ \eps2\ \est\ \Phizt + \epsl\ \st2\ \Phizz -
		\eps2\ \stt\ \Phiz = \inoi \half(\tau)\ [\ \epsl\ \Upt -
		\est\ \Upz\ ] d\tau, 					\\
					\label{eq:memphasefield2syscomp} \\
		\epsl\ \Upt - \est\ \Upz + \epsl\ \tau\ \Up = 
		\eln3\ \Phixx - 2\ \eps2\ \esx\ \Phizx + 
		\epsl\ (1 + \sx2)\ \Phizz - \eps2\ \sxx\ \Phiz\ + 	
		\epsl\ f(\Phi) + \epsl\ h,
		\end{array}
	\right.
\endeq

\nd for \( \tau \in [0,\infty) \). 
\bk

\nd The asymptotic expansion of \( \Phi \) is assumed to have the form

\begeq
	\Phi \sim \zPhi + \epsl\ \oPhi + \eps2\ \tPhi + {\cal O} (\epsl^{3}),  
	\ \ \ \ \ \ \ \ \ \ \ \ \ \ \ \ \ \mbox{as}\ \epsl 
	\rightarrow 0.
\endeq

\nd Substituting into (\ref{eq:memphasefield2syscomp}) and equating 
coefficients of the corresponding powers of \( \epsl \), we obtain the 
following problems for \( {\cal O} (1)\), \( {\cal O}(\epsl)\) and 
\( {\cal O}(\eps2) \) respectively:
\bk

\begeq
	\left\{ 
		\begin{array}{l}	
		\inoi \half(\tau)\ \est\ \zUpz\ d\tau = 0,		\\
						\label{eq:probsyscart0} \\
		\est\ \zUpz = 0,
		\end{array}
	\right.
\endeq

\begeq
	\left\{ 
		\begin{array}{l}	
		\st2\ \zPhizz = \inoi \half(\tau)\ [\ \zUpt - \est\ \oUpz\ 
		d\tau\ ],  \\
						\label{eq:probsyscart1} \\
		\zUpt - \est\ \oUpz + \tau\ \zUp = (1 + \sx2)\ \zPhizz + 
		f(\zPhi),
		\end{array}
	\right.
\endeq

\nd and

\begeq
	\left\{ 
		\begin{array}{l}	
		- 2\ \est\ \zPhizt + \st2\ \oPhizz - \stt\ \zPhiz = 
		\inoi \half(\tau)\ [\ \oUpt - \est\ \tUpz\ ]\ d\tau,	\\
						\label{eq:probsyscart2} \\
		\oUpt - \est\ \tUpz + \tau\ \oUp = - 2\ \esx\ \zPhizx + 
		(1 + \sx2)\ \oPhizz - \sxx\ \zPhiz + f'(\zPhi)\ \oPhi + h,
		\end{array}
	\right.
\endeq

\nd for all \( \tau \in [0,\infty) \). From (\ref{eq:outsideint0}) and 
(\ref{eq:outsideint1}) we have

\begeq
	\lim_{z \rightarrow \infty}\ \zUp(x,z,t;\tau) = 0,  
							\label{eq:assumpint0}
\endeq	

\nd and 

\begeq
	\lim_{z \rightarrow \infty}\ \oUp(x,z,t;\tau) = 0.
							\label{eq:assumpint1}
\endeq	

\nd From (\ref{eq:probsyscart0}) we have
\( \zUpz(x,z,t;\tau) = 0 \) for all \( \tau \in [0,\infty) \).  
Integrating with respect to \( z \) and applying condition 
(\ref{eq:assumpint0}) (assuming \(\est \neq 0 \)) yields

\begeq
	\zUp \equiv 0,				\label{eq:solsyscart0}
\endeq

\nd for all \( \tau \in [0,\infty) \). Substituting (\ref{eq:solsyscart0}) into
(\ref{eq:probsyscart1}) the second equation becomes

\begeq
	- \est\ \oUpz = (1 + \sx2)\ \zPhizz + f(\zPhi).	
						\label{eq:presolsyscart1}
\endeq

\nd We see that \( \oUpz \) does not depend on \( \tau \). 
We multiply the second equation in (\ref{eq:probsyscart1}) by 
\( \alf(\tau) \) and integrate with respect to \( \tau \). We obtain

\begdis
	\st2\ \zPhizz = [\ (1 + \sx2)\ \zPhizz + f(\zPhi)\ ]\
	\inoi \half(\tau)\ d\tau
\enddis

\nd since \( \Phi \) and \( S \) do not depend on \( \tau \). We call

\begeq
	\alf = \left( \inoi \half(\tau)\ d\tau \right)^{-1},
							\label{eq:defalf}
\endeq

\nd obtaining

\begeq
	(\ \nxicart\ ) \zPhizz + f(\zPhi) = 0.		\label{eq:probcart0}
\endeq

\nd From (\ref{eq:presolsyscart1}) we have that 
\( \est\ \oUpz = - \alf\ \st2\ \zPhizz \). Integrating with respect to \( z \) 
and applying condition (\ref{eq:assumpint1}) we obtain

\begeq
	\oUp = - \alf\ \est\ \zPhiz.			\label{eq:solsyscart1}
\endeq

\nd In order to solve (\ref{eq:probcart0}) we define a new variable

\begeq
	\xi := \frac{z}{\naxicart}.			\label{eq:defxicart}
\endeq

\nd In terms of \( \xi \), equation (\ref{eq:probcart0}) reads

\begeq
	\zPhixixi + f(\zPhi) = 0,			\label{eq:probxicart0}
\endeq

\nd whose solution is \( \zPhi = \Psi(\xi) \), the unique solution of
\(\ \ \Psi'' + f(\Psi) = 0,\ \Psi(\pm \infty) = \phi^{\pm},\ \Psi(0) = 0 \). 
Thus

\begeq
	\zPhi = \zPhi\left(\frac{z}{\naxicart}\right).	\label{eq:solxicart0}
\endeq

\nd Multiplying the second equation in (\ref{eq:probsyscart2}) by 
\( \alf(\tau) \), integrating with respect to \( \tau \), replacing 
(\ref{eq:solsyscart1}) into (\ref{eq:probsyscart2}), and rearranging terms we 
obtain

\begdis
	(\ \xicart\ ) \oPhizz + f'(\zPhi)\ \oPhi = 
	( \sxx - \alf\ \stt )\zPhiz - 2\ \alf\ \est\ \zPhizt + 
\enddis

\begeq
	+ 2\ \esx\ \zPhizx - h - \alf^{2}\ \est\ 
	\left( \inoi \half(\tau)\ \tau\ d\tau \right)\ \zPhiz.
						\label{eq:precondsyscart1}
\endeq 

\nd Setting

\begeq
	\gam = \alf^{2}\ \left( \inoi \half(\tau)\ \tau\ d\tau \right). 
						\label{eq:defgama}
\endeq

\nd In terms of \( \xi \), \( x \) and \( t \) equation 
(\ref{eq:precondsyscart1}) reads (see Appendix A)

\begdis
	\oPhixixi + f'(\zPhi)\ \oPhi = \frac{\sxx - \alf\ \stt - 
	\gam \est}{\axicart}\ \zPhixi - 
	\frac{2 \esx ( \esx\ \sxx - \alf\ \est\ \sxt )}{\bxicart} 
	\ ( \xi\ \zPhixixi + \zPhixi ) +
\enddis
 
\begeq
	+ \frac{2\ \alf\ \est\ ( \esx\ \sxt -  \alf\ \est\ \stt )}{\bxicart}  
	\ ( \xi\ \zPhixixi + \zPhixi ) - h. 
						     \label{eq:precondxicart1} 
\endeq

\nd It is straightforward to check that \( \Psi'(\xi) \) satisfies 
the homogeneous equation 

\begeq
	 \oPhixixi + f'(\zPhi)\ \oPhi = 0.		\label{eq:homog}
\endeq

\nd This implies that the operator \( \Lambda \) defined by

\begeq
	\Lambda := \frac{\partial^{2}}{\partial \xi^{2}} + f'(\Psi'(\xi))
							\label{eq:operat}
\endeq

\nd has a simple eigenvalue at the origin with \( \Psi' \) as the 
corresponding eigenfunction. Now the solvability condition for equation 
(\ref{eq:precondxicart1}) gives

\begdis
	\frac{\sxx - \alf\ \stt -  \gam\ \est}{\axicart} 
	\int_{-\infty}^{\infty} (\Psi')^{2} d\xi -
	h \int_{-\infty}^{\infty} \Psi' d\xi +
\enddis

\begeq
	- \frac{ 2\ \esx ( \esx\ \sxx - \alf\ \est\ \sxt ) -
	2\ \alf\ \est\ ( \esx\ \sxt - \alf\ \est\ \stt ) }{\bxicart}
	\int_{-\infty}^{\infty} (\xi\ \Psi'' + \Psi' ) \Psi' d\xi = 0.
							\label{eq:condcart}
\endeq

\nd A simple calculation shows that

\begeq
	\int_{-\infty}^{\infty} \xi \Psi' \Psi'' d\xi = - \frac{1}{2} 
	\int_{-\infty}^{\infty} (\Psi')^{2} d\xi.	\label{eq:integ}
\endeq

\nd Defining

\begeq
	\nu := h\ \frac{\Psi(+\infty) - \Psi(-\infty)}{
	\int_{-\infty}^{\infty} (\Psi')^{2} d\xi}, 	    \label{eq:defpsi}
\endeq

\nd Substituting (\ref{eq:integ}) and (\ref{eq:defpsi}) into 
(\ref{eq:condcart}),
multiplying equation (\ref{eq:condcart}) by \( \bxicart \) and rearranging 
terms we obtain (\ref{eq:mbornexcart}).
Note that for \( f(\phi) = (\phi - \phi^{3}) / 2 \) 
(Ginzburg-Landau theory), \( \Psi(\xi) = \tanh \frac{\xi}{2} \) and
\( \nu := 3\ h \), whereas for \( f(\phi) = \sin \phi \), 
\( \Psi(\xi) = 4\ \tan^{-1} e^{\xi} - \pi \) and \( \nu := \frac{\pi}{4}\ h \).

\section{The phase field equations with memory - asymptotic analysis in local
coordinates}

\subsection{Assumptions and definitions}

\nd For the asymptotic analysis, using  local 
coordinates, we set \( x=(x_{1},x_{2}) \) and call \( d(x) \) the distance 
from \( x \) to \( \Gamma \); i.e., \( d(x) = dist(x,\Gamma) \). We next 
define a local orthogonal coordinate system \( (r,s) \) in a neighborhood of 
\( \Gamma \) in the following way

\begeq
	r(x,t;\epsilon) = 
	\left\{ \begin{array}{lll}
		\ \ d(x)  & if & \phi(x) > 0   \\
		                          &   &   \label{eq:defas33}    \\
		-d(x) & if & \phi(x) < 0,
		\end{array}
	\right.
\endeq
\bk

\nd and \( s(x,t;\epsilon) \), a smooth function of \( t \), such that on
\( \Gamma(t;\epsilon) \) it measures arclength from some point which  
moves normally to \( \Gamma \) as \( t \) varies. The assumed initial 
smoothness of \( \Gamma(t;\epsilon) \) implies that \( r \) is a smooth 
function, at least, in a sufficiently small neighborhood of \( \Gamma \).
\bk  

\nd The outer expansions of \( \phi \) and \( u \) are assumed to have the form

\begeq
	\phi = \phi(x,t;\epsl) = \zphi(x,t) + \epsl\ \ophi(x,t) + 
	\eps2\ \tphi(x,t) + {\cal O}(\epsl^{3})
							\label{eq:asymexp1}
\endeq

\nd and

\begeq
	u = u(x,t;\epsl) = \zu(x,t) + \epsl\ \ou(x,t) + \eps2\ u^{2}(x,t) +
	{\cal O}(\epsl^{3}).
							\label{eq:uasymexp1}
\endeq

\nd In order to determine the inner expansions we first define the inner 
variable

\begeq
	z(x,t;\epsilon) := \frac{r(x,t;\epsilon)}{\epsilon} 
							\label{eq:asymexp2}
\endeq

\noindent and then assume the inner expansions to be given by

\begeq
	\phi = \Phi(z,s,t;\epsl) = \zPhi(z,s,t) + \epsl\ \oPhi(z,s,t) + 
	{\cal O}(\eps2)
							\label{eq:asymexp3}
\endeq 

\nd and

\begeq
	u = U(z,s,t;\epsl) = \zU(z,s,t) + \epsl\ \oU(z,s,t) + 
	{\cal O}(\eps2).
							\label{eq:uasymexp3}
\endeq 

\nd The very definition of \( \Gamma \) requires \( \Phi(0,s,t;\epsilon) 
= 0 \). In what follows we will use the following notation to refer to 
any variable \( g \) evaluated by approaching \( \Gamma \) from either 
side (\( r > 0 \) or \( r < 0 \)):

\begeq
	g \mid_{\Gamma^{\pm}} = \lim_{r \rightarrow 0^{\pm}} 
	g(r,s,t;\epsilon),
							\label{eq:asymexp4}
\endeq

\begeq
	g_{r} \mid_{\Gamma^{\pm}} = \lim_{r \rightarrow 0^{\pm}} 
		g_{r}(r,s,t;\epsilon).
							\label{eq:asymexp5}
\endeq

\nd The following relations between the inner and outer variables obtained in 
\cite{kn:fif1} are assumed to hold  as \( \rho \rightarrow \pm \infty \).

\begeq
	G^{0}(\rho,s,t) = g^{0}(0^{\pm},s,t),		\label{eq:matchcon0}
\endeq

\begeq
	G^{1}(\rho,s,t) = g^{1}(0^{\pm},s,t) + \rho\, g^{0}_{r}(0^{\pm},s,t).
							\label{eq:matchcon1}
\endeq

\subsection{Derivation of the equations of motion}

\nd Let us look at the system (\ref{eq:defas2}) assuming that \( u = 0 \) 
initially and along the boundary. 

\subsubsection{Outer problems}

\nd Substituting  (\ref{eq:asymexp1}) and (\ref{eq:uasymexp1}) into 
(\ref{eq:defas2}) and equating coefficients of the corresponding 
powers of \( \epsl \) we obtain the \( {\cal O}(1) \) and \( {\cal O}(\epsl) \)
outer problems respectively for points where the interface has not yet arrived:

\begeq
	\left\{\begin{array}{l}
		\zut = a_{1}\ \ast\ \zlapu, 	\\
							\\
		a_{2} \ast\ f(\zphi) = 0,				
							\label{eq:uoutprob0}
		\end{array}
	\right.
\endeq

\nd and

\begeq
	\left\{\begin{array}{l}
		\out = a_{1}\ \ast\ \olapu, 	\\
							\\
		a_{2}\ \ast\ [ f'(\zphi)\ \ophi + \zu\ ] = 0. 	  
							\label{eq:uoutprob1}
		\end{array}
	\right.
\endeq

\nd The solution of (\ref{eq:uoutprob0}), given the assumed initially and 
boundary conditions for \( u \), is \( \zu \equiv 0 \), \( \zphi = \pm 1 \).
The solution of (\ref{eq:uoutprob1}) is \( \ou \equiv 0 \),
\( \ophi \equiv 0 \). 

\subsubsection{Inner problems}

\nd From the solution of the outer 
problems (\ref{eq:uoutprob0}) and (\ref{eq:uoutprob1}) and  for points where 
the moving front has not yet arrived, we have 

\begeq
	\chi^{0}(t;w) = 0,			
	\ \ \ \ \ \ \ \ \ \ \mbox{and}\ \ \ \ \ \ \ \ \ \
	v^{0}(t;w) = 0,			\label{eq:poutsideint0}\
\endeq

\nd and

\begeq
	\chi^{1}(t;w) = 0.	
	\ \ \ \ \ \ \ \ \ \ \mbox{and}\ \ \ \ \ \ \ \ \ \
	v^{1}(t;w) = 0,				\label{eq:poutsideint1}
\endeq

\nd for all \( x \in \Omega \) and \( w \in (-\infty,\infty) \).
\bk

\nd System (\ref{eq:defas2fou}) expressed
in the \( (z,s,t) \) coordinates (see appendix B), after differentiating the 
first and third equations with respect to \( t \) and calling 
\( v = V(z,s,t;w,\epsl) \) and \( \chi = \Up(z,s,t;w,\epsl) \), becomes

\begeq
	\left\{ 
		\begin{array}{l}	
		\radt2\ \Uzz + \epsl\ [\ 2\ \ardt\ \Uzt + \radtt\ \Uz\ ] +
		\eps2\ [ \Utt + 2\ \ract\ \Ust\ + \arct2\ \Uss + \arctt\ \Us +
		2\ \ardt\ \ract\ \Uzs + \radt2\ \Phizz ] = 		\\
									\\
		\ \ \ \ \ \ \ \ \ \ \ \ \ \ \ \ \ \ \ \ \ \ \ \ \ 
		\ \ \ \ \ \ \ \ \ \ \ \ \ \ \ \ \ \ \ \ \ \ \ \ \
		= \inii \ohalf(w)\ [ \eps2\ \Vt + \epsl\ \ardt\ \Vz + 
		\eps2 \ract\ \Vs ]\ dw + {\cal O}(\eln3), 		\\
									\\
		\epsl\ \ardt\ \Vz + \eps2\ [ \Vt + \ract\ \Vs + i\ w\ V ] =
		\Uzz + \epsl\ \laprad\ \Uz + \eps2\ [\ \Uss | \nabla s | + \Us 
		\Delta s\ ],						\\
									\\
		\epsl\ \radt2\ \Phizz + \eps2\ [\ 2\ \ardt\ \Phizt + 
		\radtt\ \Phiz + 2\ \ardt\ \ract\ \Phizs\ ] = 
		\inii \thalf(w) [\ \epsl\ \Upt + \ardt\ \Upz + 
		\epsl\ \ract\ \Ups\ ]\ dw + {\cal O}(\eln3), 	\\
									\\
		\ardt\ \Upz + \epsl\ [\ \Upt\ + \ract\ \Ups\ + i\ w\ 
		\Upsilon\ ] = \epsl\ [\ \Phizz + f(\Phi)\ ] + 
		\eps2\ [\ \laprad\ \Phiz + U\ ] + {\cal O}(\eln3),	 
							\label{eq:asymexpcomp}
		\end{array}
	\right.
\endeq
\bk

\nd for \( w \in (-\infty,\infty) \). From (\ref{eq:poutsideint0}) and 
(\ref{eq:poutsideint1}) we have

\begeq
	\lim_{z \rightarrow \infty}\ \zUp(x,z,t;w) = 0,
	\ \ \ \ \ \ \ \ \ \ \mbox{and}\ \ \ \ \ \ \ \ \ \ 
 	\lim_{z \rightarrow \infty}\ V^{0}(x,z,t;w) = 0,
							\label{eq:passumpint0}
\endeq	

\nd and 

\begeq
	\lim_{z \rightarrow \infty}\ \oUp(x,z,t;w) = 0,
	\ \ \ \ \ \ \ \ \ \ \mbox{and}\ \ \ \ \ \ \ \ \ \
	\lim_{z \rightarrow \infty}\ V^{1}(x,z,t;w) = 0.
							\label{eq:passumpint1}
\endeq	

\nd Substituting (\ref{eq:asymexp3}) and 
(\ref{eq:uasymexp3}) into (\ref{eq:asymexpcomp}) and equating coefficients of 
the corresponding powers of \( \epsl\ \) we obtain the \( {\cal O} (1)\), 
\( {\cal O}(\epsl)\) and \( {\cal O}(\eps2) \) problems respectively. 
\bk

\nd \underline{\( {\cal O} (1)\)}:

\begeq
	\left\{ 
		\begin{array}{l}	
		\radt2\ \zUzz = 0,					\\
									\\
		\zUzz = 0,						\\
									\\
		\inii \thalf(w)\ \ardt\ \zUpz dw = 0,		\\
									\\
		\ardt\ \zUpz = 0,			\label{eq:inasympcomp0}
		\end{array}
	\right.
\endeq

\nd \underline{\( {\cal O} (\epsl)\)}:

\begeq
	\left\{ 
		\begin{array}{l}	
		2\ \ardt\ \zUzt + \radt2\ \oUzz\ \radtt\ \zUz = 
		\inii \ohalf(w)\ \ardt\ \zVz\ dw = 0, \\
									\\
		\ardt\ \zVz = \oUzz + \laprad\ \zUz,			   \\
									\\
		\radt2\ \zPhizz = \inii \thalf(w)\ [\ \zUpt + 
		\ardt\ \oUpz + \ract\ \zUps\ ] dw,			\\
									\\
		\zUpt + \ardt\ \oUpz + \ract\ \zUps\ + i\ w\ \zUp =
		\zPhizz + f(\zPhi),			\label{eq:inasympcomp1}
		\end{array}
	\right.
\endeq

\newpage

\nd \underline{\( {\cal O} (\eps2)\)}:

\begeq
	\left\{ 
		\begin{array}{l}	
		\radt2\ \tUzz + 2\ \ardt\ \oUzt + \radtt\ \oUz + \zUtt + 
		2\ \ract\ \zUst + \arct2\ \zUss + \arctt\ \zUs + 
		2\ \ardt\ \ract\ \zUzs 					\\
									\\
		\ \ \ \ \ \ \ \ \ \ \ \ \ \ \ \ \ \ \ \ \ \ \ \ \
		\ \ \ \ \ \ \ \ \ \ \ \
		+ \radt2\ \zPhizz = \inii \ohalf(w) 
		[\ \zVt + \ardt\ \oVz + \ract\ \zVs\ ]\ dw, 		\\
									\\
		\ardt\ \oVz + \zVt + \ract\ \zVs + i\ w\ \zV = 
		\tUzz + \laprad\ \oUz + \zUss | \nabla s | + \zUs\ \Delta s, \\
									\\
		\radt2\ \oPhizz + 2\ \ardt\ \zPhizt + \radtt\ \zPhiz + 
		2\ \ardt\ \ract\ \zPhizs = \inii \thalf(w) [\ \oUpt + 
		\ardt\ \tUpz + \ract\ \oUps\ ]\ dw, 			   \\
									\\
		\oUpt + \ardt\ \tUpz + \ract\ \oUps + i\ w\ \oUp = 
		\oPhizz + f'(\zPhi)\ \oPhi + \laprad\ \zPhiz + \zU,  	   
							\label{eq:inasympcomp2}
		\end{array}
	\right.
\endeq
\bk

\nd for \( w \in (-\infty,\infty) \). We call

\begeq
	\alf_{i} = \left( \inii \half_{i}(w)\ dw \right)^{-1},
							\label{eq:defalfi}
\endeq

\nd for \( i = 1, 2 \).
A bounded solution of the first two equations in (\ref{eq:inasympcomp0})
satisfying the matching conditions (\ref{eq:matchcon0}) is \( \oU \equiv 0 \). 
From the third and fourth equations in (\ref{eq:inasympcomp0}) we have
\( \zUpz(x,z,t;w) = 0 \) for \( w \in (-\infty,\infty) \).  
Integrating with respect to \( z \) (assuming that \( r_{t} \neq 0 \)) and 
applying condition (\ref{eq:passumpint0}) yields

\begeq
	\zUp \equiv 0,				\label{eq:solsysloc0}
\endeq

\nd for \( w \in (-\infty,\infty) \). Substituting \( \oU \equiv 0 \) in the 
first two equations in (\ref{eq:inasympcomp1}), multiplying the second and 
fourth equations in (\ref{eq:inasympcomp1}) by \( \ohalf \) and \( \thalf \) 
respectively, and integrating with respect to \( w \) yields

\begeq
	\left\{\begin{array}{l}
		( \oalf - \radt2 )\ \oUzz = 0, 				\\
									\\
		( \dximove )\ \zPhizz + f(\zphi) = 0,	\label{eq:uinprob0}
		\end{array}
	\right.
\endeq

\nd Assuming that \( \radt2 \neq \oalf \), the bounded solution of the first 
equation in (\ref{eq:uinprob0}) satisfying the matching conditions 
(\ref{eq:matchcon0}) is \( \oU \equiv 0 \). To solve the second equation in 
(\ref{eq:uinprob0}), we assume that \( \radt2 \neq \talf \) and
we define the new variable  

\begeq
	\xi := \frac{z}{\adximove}.			\label{eq:uinsol00}
\endeq

\nd In terms of \( (\xi,s,t) \), equation the second equation in 
(\ref{eq:uinprob0}) reads

\begeq
	\zPhixixi + f(\zPhi) = 0,			\label{eq:uinsol01}
\endeq

\nd whose solution is \( \zPhi = \Psi(\xi) \), the unique solution of
\(\ \ \Psi'' + f(\Psi) = 0,\ \Psi(\pm \infty) = \pm 1,\ \psi(0) = 0 \). Thus

\begeq
	\zPhi = \zPhi\left(\frac{z}{\adximove}\right),	\label{eq:uinsol02}
\endeq

\nd which satisfies (\ref{eq:matchcon0}). From the fourth equation in 
(\ref{eq:inasympcomp1}) we have \( \ardt\ \oUpz = \talf\ \radt2\ \zPhizz \).
Integrating with respect to \( z \) and applying condition 
(\ref{eq:passumpint1}) we get

\begeq
	\oUp = - \talf\ \radt2\ \zPhiz.			\label{eq:solsyscart1l}
\endeq 

\nd Substituting (\ref{eq:uinsol02}) into the third and fourth equations in 
(\ref{eq:inasympcomp2}), multiplying the fourth equation by \( \thalf \) and
integrating with respect to \( w \) yields

\begdis
	( \dximove )\ \oPhizz + f'(\zPhi)\ \oPhi = 
\enddis

\begeq
	= ( \talf\ \radtt + \tgam\ \talf\ \ardt - \laprad )\ \zPhiz + 
	+ 2\  \ardt\ \zPhizt, 				\label{eq:afuinprob1}
\endeq

\nd where 

\begeq
	\tgam = i\ \talf^{2}\ \left( \inii \thalf(w)\ w\ dw \right).
						\label{eq:defgama2}
\endeq

\nd Equation (\ref{eq:afuinprob1}) expressed in the (\(\xi,s,t\)) coordinate 
system reads (see appendix B)

\begdis
	\oPhixixi + f'(\zPhi)\ \oPhi = 
\enddis

\begeq
	\frac{2\ \radt2\ \radtt}{\bdximove}\ ( \xi\ \zPhixixi + \zPhixi )
	 + \frac{\radtt + \tgam\ \ardt - \laprad}{\adximove}\ \zPhixi.
							\label{eq:uinsol11ph}
\endeq

\nd It is straightforward to check that \( \Psi'(\xi) \) satisfies 
the homogeneous equation 

\nd \( \oPhixixi + f'(\zPhi)\ \oPhi = 0 \). That means 
that the operator \( \Lambda := \frac{\partial^{2}}{\partial \xi^{2}} + 
f'(\Psi'(\xi)) \) has a simple eigenvalue at the origin with \( \Psi' \) as 
the corresponding eigenfuction. The solvability condition for the 
equation (\ref{eq:uinsol11ph}) now gives

\begdis
	\frac{2\ \radt2\ \radtt}{\bdximove}\ \int_{-\infty}^{\infty} 
	(\xi\ \Psi'' + \Psi' )\ \Psi'\ d\xi + 
\enddis

\begeq
	\frac{\radtt + \tgam\ \ardt
	 - \laprad}{\adximove}\ \int_{-\infty}^{\infty} (\Psi')^{2} d\xi = 0.
							\label{eq:uinsol12ph}
\endeq

\nd A simple calculation shows that \( \int_{-\infty}^{\infty} \xi \Psi' 
\Psi'' d\xi = - \frac{1}{2} \int_{-\infty}^{\infty} (\Psi')^{2} d\xi \).
Hence multiplying equation (\ref{eq:uinsol12ph}) by \( \bdximove\) and
rearranging terms one obtains

\begeq
	\frac{\radtt}{\dximove} + \tgam\ \ardt = \kap.
							\label{eq:flowmcmemp}
\endeq

\nd Taking into consideration that on the interface \( \laprad = \kap \),
the curvature of the interface, and that \( \ardt = - v \), its normal 
velocity \cite{kn:fifpen1}, equation (\ref{eq:flowmcmemp}) becomes
(\ref{eq:flowmcmemph}).
\bk

\section{Conclusions}

\nd In this paper we showed that to the leading order and for a large class
of kernels \( a \), \( a_{1} \) and \( a_{2} \)  under suitable 
assumptions on them, the law governing the
evolution of interfaces for the integro-differential equation 
(\ref{eq:memphasefield22}) is the same as for the differential equation
(\ref{eq:phasefield22}). It is easy to see that \( \tgam \) is given by 

\begdis
	 \gam = -\frac{a'(0)}{(a(0))^{2}}.
\enddis 
\bk

\nd For equation (\ref{eq:defas2}) the result is similar with 
\( \gam \) and \( a \) changed by \( \tgam \) and \( a_{2} \).  Our 
derivation is only valid for advancing fronts; i.e., non-oscillating 
fronts in the sense that there exist points on the interface with vanishing 
velocity. This restriction prevent us from analyzing system where 
those oscillations may be relevant. This assumption is not necessary 
when the kernels are exponentially decreasing \cite{kn:rotnep1,kn:rotnep2}. 
\bk

\nd To solve the \( {\cal O} (\epsl) \) problem we have assumed that 
\( |\ardt|  \neq \sqrt{\oalf} \) (\( \oalf \) being similar to \( \talf \)). 
On the other hand, assuming that \( 1 - \talf\ v^{2} > 0 \) initially and that 
there is not change of sign in this direction during the evolution, a fact 
which is true for the circular case, we can easily see from equation 
(\ref{eq:flowmcmemph}) that \( |\ardt| < \frac{1}{\sqrt{\talf}} \). Therefore,
if \( \oalf \talf \geq 1 \) the former assumption does not add further 
restrictions on the interfacial motion. 
\bk

\nd \underline{Acknowledgement}: The authors are indebted to A. Novick-Cohen
for the formulation of the problem and to Ch. Charach for valuable discussions.
\bk

\appendix

\section{Cartesian coordinates}

\nd \underline{{\em From Cartesian to (\(z,x,t\)) coordinates}}:

\nd To go from Cartesian to (\(z,x,t\)) coordinates we transform 
derivatives as follows

\begdis
	\phit = \Phit - \invepsl\ \est\ \Phiz,
\enddis

\begdis
	\phitt = \Phitt - \invepsl\ 2\ \est\ \Phizt + \inveps2\ \st2\ \Phizz -
	\invepsl\ \stt\ \Phiz,
\enddis

\begdis
	\phixx = \Phixx - \invepsl\ 2\ \esx\ \Phizx + \inveps2\ \sx2\ \Phizz -
	\invepsl\ \sxx\ \Phiz,
\enddis

\begdis
	\phiyy = \inveps2 \Phizz,
\enddis

\begdis
	\phitxx = \Phitxx - \invepsl\ \est\ \Phizxx - \invepsl\ 2\ \sxt\ 
	\Phizx - \invepsl\ 2\ \esx\ \Phitxz + \inveps2\ 2 \esx\ \est\ \Phixzz 
	+ 
\enddis

\begdis
	+ \inveps2\ 2\ \esx\ \sxt\ \Phizz + \inveps2\ \sx2\ \Phitzz - 
	\inv3eps\ \sx2\ \est\ \Phizzz - \invepsl\ \tsxx\ \Phiz -
	\invepsl\ \sxx\ \Phizt + \inveps2\ \sxx\ \est\ \Phizz,
\enddis

\begdis
	\phityy = \inveps2\ \Phitzz - \inv3eps\ \est\ \Phizzz.
\enddis 


%

\nd \underline{{\em From (\(z,x,t\)) to (\(\xi,x,t\)) coordinates}}:

\nd To go from (\(z,x,t\)) to (\(\xi,x,t\)) coordinates, derivatives are
transformed as follows

\begdis
	\xiz = \invaxicart,
\enddis

\begdis
	\xix = - z \invbxicart\ ( \esx \sxx - \alf \est \sxt ),
\enddis
 
\begdis
	\xit = - z \invbxicart\ ( \esx \sxt - \alf \est \stt ),
\enddis

\begdis
	\xizx = - \invbxicart\ ( \esx \sxx - \alf \est \sxt ),
\enddis

\begdis
	\xizt = - \invbxicart\ ( \esx \sxt - \alf \est \stt ).
\enddis

\begeq
	\zPhiz = \zPhixi \xiz = \frac{1}{\axicart} \zPhixi		
						\label{eq:changecart1}
\endeq

\begeq
	\zPhizt = \zPhixixi \xiz \xit + \zPhixi \xizt = 
	- \frac{\esx \sxt - \alf \est \stt}{\bxicart}\ ( \xi \zPhixixi + 
	\zPhixi ).					\label{eq:changecart2}
\endeq

\begeq
	\zPhizx = \zPhixixi \xiz \xix + \zPhixi \xizx = 
	- \frac{\esx \sxx - \alf \est \sxt}{\bxicart}\ ( \xi \zPhixixi + 
	\zPhixi ).					\label{eq:changecart3}
\endeq

\section{Local coordinates}

\subsection{From Cartesian to (\(r,s,t\)) and \( (z,s,t) \) 
coordinates}:

\nd To go from Cartesian to \( (r,s,t) \) coordinates we transform derivatives
as follows ( using the fact that \( \left| \nabla r \right| \equiv 1 \) ) 

\begdis
	\phit = \Phit + \Phir \ardt + \Phis \ract,
\enddis

\begdis 
	\phitt = \Phitt + 2 \Phirt \ardt + 2 \Phist \ract + \Phirr \radt2 
	+ \Phiss \arct2 + \Phir \radtt + \Phis \arctt + 2 \Phirs \ardt \ract,
\enddis

\nd and

\begdis
	\lap-phi = \Phirr + \Phir \laprad + \Phiss \gradarc2 + \Phis \laparc,
\enddis




\noindent where the operators \( \nabla \) and \( \Delta \) refers only to the 
spatial variable \( x \). These derivatives expressed in terms of \( (z,s,t) \)
are

\begdis
	\phit = \Phit + \invepsl \Phiz \ardt + \Phis \ract,
\enddis

\begdis 
	\phitt = \Phitt + 2 \invepsl \Phizt \ardt + 2 \Phist \ract + 
	\inveps2 \Phizz \radt2 + \Phiss \arct2 + \invepsl \Phiz \radtt + 
	\Phis \arctt + 2 \invepsl \Phizs \ardt \ract,
\enddis

\nd and

\begdis
	\lap-phi = \inveps2 \Phizz + \invepsl \Phiz \laprad + 
	\Phiss \gradarc2 + \Phis \laparc,
\enddis




\subsection{\em From (\(z,s,t\)) to (\(\xi,s,t\)) coordinates}

\nd To go from (\(z,s,t\)) to (\(\xi,s,t\)) coordinates, derivatives are
transformed as follows

\begdis
	\xiz = \tinvaximove,
\enddis

\begdis
	\xit = z \tinvbximove\ \alf \ardt \radtt
\enddis

\begdis
	\xizt = \tinvbximove\ \alf \ardt \radtt.
\enddis

\begeq
	\zPhiz = \zPhixi \xiz = \frac{1}{\taximove} \zPhixi	
						\label{eq:changemove1}
\endeq

\begeq
	\zPhizt = \zPhixixi \xiz \xit + \zPhixi \xizt = 
	\frac{\alf \ardt \radtt }{\tbximove}\ ( \xi \zPhixixi + 
	\zPhixi ).					\label{eq:changemove2}
\endeq

\bibliographystyle{unsrt}
\bibliography{tesis}

\end{document}